\documentclass[10pt,english,aps]{revtex4-1}
\usepackage[latin9]{inputenc}
\usepackage{textcomp}
\usepackage{amsmath}
\usepackage{amssymb}
\usepackage{graphicx}

\makeatletter
\@ifundefined{textcolor}{}
{%
 \definecolor{BLACK}{gray}{0}
 \definecolor{WHITE}{gray}{1}
 \definecolor{RED}{rgb}{1,0,0}
 \definecolor{GREEN}{rgb}{0,1,0}
 \definecolor{BLUE}{rgb}{0,0,1}
 \definecolor{CYAN}{cmyk}{1,0,0,0}
 \definecolor{MAGENTA}{cmyk}{0,1,0,0}
 \definecolor{YELLOW}{cmyk}{0,0,1,0}
}

\begin{document}

\title{Characterization of spatiotemporal chaos in Kerr optical frequency {comb generators} and in fiber cavity}
\author{Z. Liu$^{1}$, M. Ouali$^{1}$,  S. Coulibaly$^{1}$, M.G. Clerc$^{2}$,  M. Taki$^{1}$ and M. Tlidi$^{3}$}
\address{Universit\'e de Lille, CNRS, UMR 8523 - PhLAM - Physique des Lasers 
Atomes et Mol\'ecules, F-59000 Lille, France$^{1}$\\
Departamento de F\'{i}sica, FCFM, Universidad de Chile, Casilla 487-3, Santiago, Chile$^{2}$\\
Universit\'{e} Libre de Bruxelles (U.L.B.), Facult{\'{e}} des Sciences,  CP. 231, Campus Plaine, B-1050 Bruxelles, Belgium$^{3}$
}

\begin{abstract}
Complex spatiotemporal dynamics have been a subject of recent experimental 
investigations in  optical frequency comb microresonators and  in driven fiber cavities 
with a Kerr-type media. We show that this complex behavior has a spatiotemporal 
chaotic nature. We  determine numerically the Lyapunov spectra, allowing to 
characterize different dynamical behavior occurring in these simple devices. The 
Yorke-Kaplan dimension is used as an order parameter to characterize the bifurcation 
diagram. We identify a wide regime of parameters where 
the system exhibits a coexistence between the spatiotemporal chaos, the oscillatory 
localized structure, and the homogeneous steady state. 
The destabilization of an oscillatory localized state 
through radiation of counter propagative fronts between the homogeneous and the 
spatiotemporal chaotic states is analyzed. To characterize better the spatiotemporal 
chaos, we estimate the front speed as a function of the pump intensity.
\end{abstract}

\maketitle

Experiments supported by numerical simulations of driven cavities such as whispering-gallery-mode microresonators leading to optical frequency comb generation have demonstrated  the existence of complex spatiotemporal dynamics \cite{Chembo_10}. Similar complex 
dynamics have been observed in all fiber cavity \cite{Mitschke96,Mitschke96b,Coen_Optica_15}.
 In most of  these studies, complex behaviors are characterized by a power spectrum \cite{Chembo_10},
 filtering spatiotemporal diagrams \cite{Coen_Optica_15}, embedding dimension, and
time series analysis \cite{Mitschke96,Mitschke96b}. However, these tools are 
 inadequate to distinguish between  spatiotemporal chaos, low dimensional chaos, and turbulence. 
 A classification of  these phenomena has been reported in the literature (see for instance \cite{Ruelle,Manneville90,Pikovsky,Nicolis,Marcel2013,SSM2016}). In the case of spatiotemporal chaos, the Lyapunov spectrum has a continuous set of  positive values. This matches the definition that has been proposed in \cite{Manneville90,Pikovsky}. In the case of a low dimensional chaos, the Lyapunov spectrum 
 possesses a discrete set of positive values. 
However, the turbulence or weak turbulence are characterized by a power law cascade of a scalar quantity such as energy, norm, etc  \cite{Frisch}. On the basis of the Lyapunov spectrum, we cannot conclude that the system develops a turbulence.

In this letter, we characterize the complex behavior reported in the paradigmatic  Lugiato-Lefever equation (LLE, \cite{Lugiato1987}) that  describes  Kerr optical frequency combs  and  fiber cavities. 
For this purpose, we use a rigorous tools of dynamical systems theory.
We show that this complex behavior has a spatiotemporal chaotic nature. 
We estimate the Lyapunov spectra. The Yorke-Kaplan dimension ($D_{YK}$) is used as an order 
parameter  to establish the bifurcation diagram of the
spatiotemporal chaos. In addition, we show that the spatiotemporal chaos, the oscillatory localized state and the homogeneous steady state (HSS) can coexist in a finite range of the pumping intensity. The destabilization 
of an oscillatory localized state through radiation of counter propagative fronts between 
the HSS and the spatiotemporal chaotic state is also discussed by estimating the front speed 
as a function of the pump intensity.

Driven Kerr cavities with a high Fresnel number assuming that the cavity is much shorter than the diffraction and the nonlinearity spatial scales---is described  in the mean field limit by the LLE \cite{Lugiato1987}. 
This equation has been extended to model both  
fiber cavities \cite{Haelterman92,Haelterman92a} and optical frequency combs generation 
\cite{Chembo_13,Wabnitz2013,Chembo2014}, in which the diffraction is replaced by dispersion. This model reads
\begin{align}
\frac{\partial \psi}{\partial t}=S-(\alpha+i\delta)
\psi+\frac{i}{2}\frac{\partial^{2} \psi}{\partial \tau^{2}}+i\vert\psi\vert^{2}\psi,
\label{Eq-DDNLS}
\end{align}
where $\psi(t,\tau)$ is the normalized slowly varying envelope of the electric field that circulates within 
the cavity and $S$ is the amplitude of the injected field which is real and constant. 
The time variable $t$ corresponds to the slow evolution 
of $\psi$ over successive round-trips. 
$\tau$ accounts for the fast dynamics that describes how the electric field envelop changes along the fiber \cite{Haelterman92,Haelterman92a,Chembo_13}. The parameters $\alpha$ and $\delta$ are the cavity losses, and the cavity detuning, respectively.
In addition, Eq.~(\ref{Eq-DDNLS})  has been derived in the context of left-handed materials \cite{Kockaert_2006}. 
Note that, Eq.~(\ref{Eq-DDNLS}) has been derived in  early reports to describe the  plasma driven by an radio frequency field \cite{DDNLS-Plasma,DDNLS} 
and the condensate in the presence of an applied ac field \cite{DDNLS-Condensate}. 
%{\bf It is well known a localized state of Eq.~(\ref{Eq-DDNLS}), in the limit of small  $S$ and $\alpha$, exhibits an oscillatory behavior  as result of  Andronov-Hopf bifurcation \cite{Nozaki1985,Nozaki1986,Taki1989}.}
\begin{figure}[t]
\centering
\includegraphics[width=8cm]{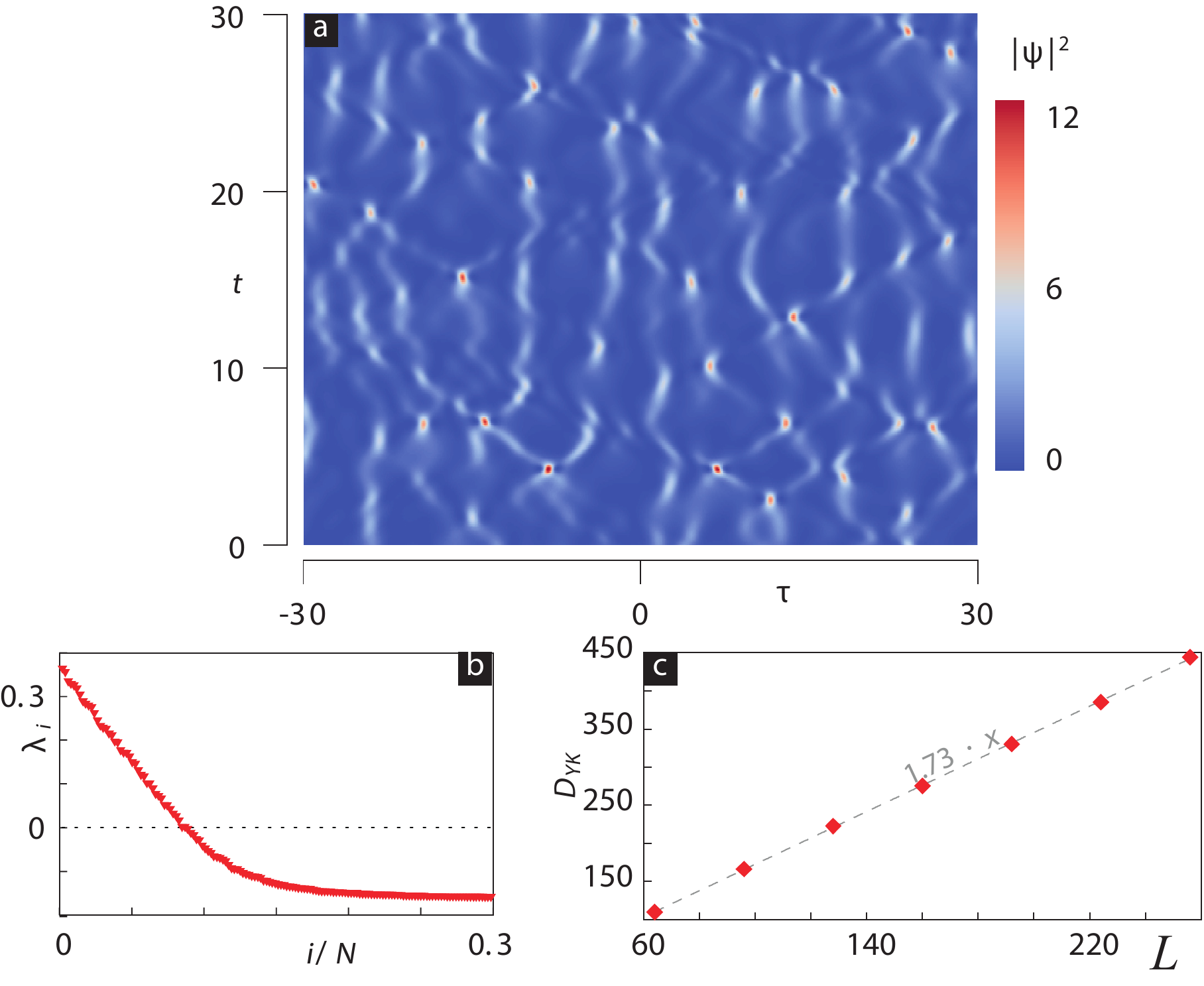}
\caption{ (color online) Spatiotemporal chaos.
(a) the $\tau$-$t$ map shows a complex spatiotemporal behavior 
obtained by numerical simulation of  Eq.~(\ref{Eq-DDNLS}) with $\alpha=0.16$,
$\delta=1$, and $S^2=0.16$ with 512 grid points. (b) The corresponding 
 Lyapunov spectrum,  and (c) the Yorke-Kaplan dimension as 
function of the system size $L$ 
is indicated by diamond red points. $L=512 \Delta \tau$  with $\Delta \tau$ is the step-size integration.
The linear growth 
of $D_{YK}$ dimension is fitted by a slope of 1.73 as shown by a gray dashed line.}
\label{Fig-Cavity_figure} 
\end{figure}
\begin{figure}[tb]
\centering 
\includegraphics[width=8.0cm] {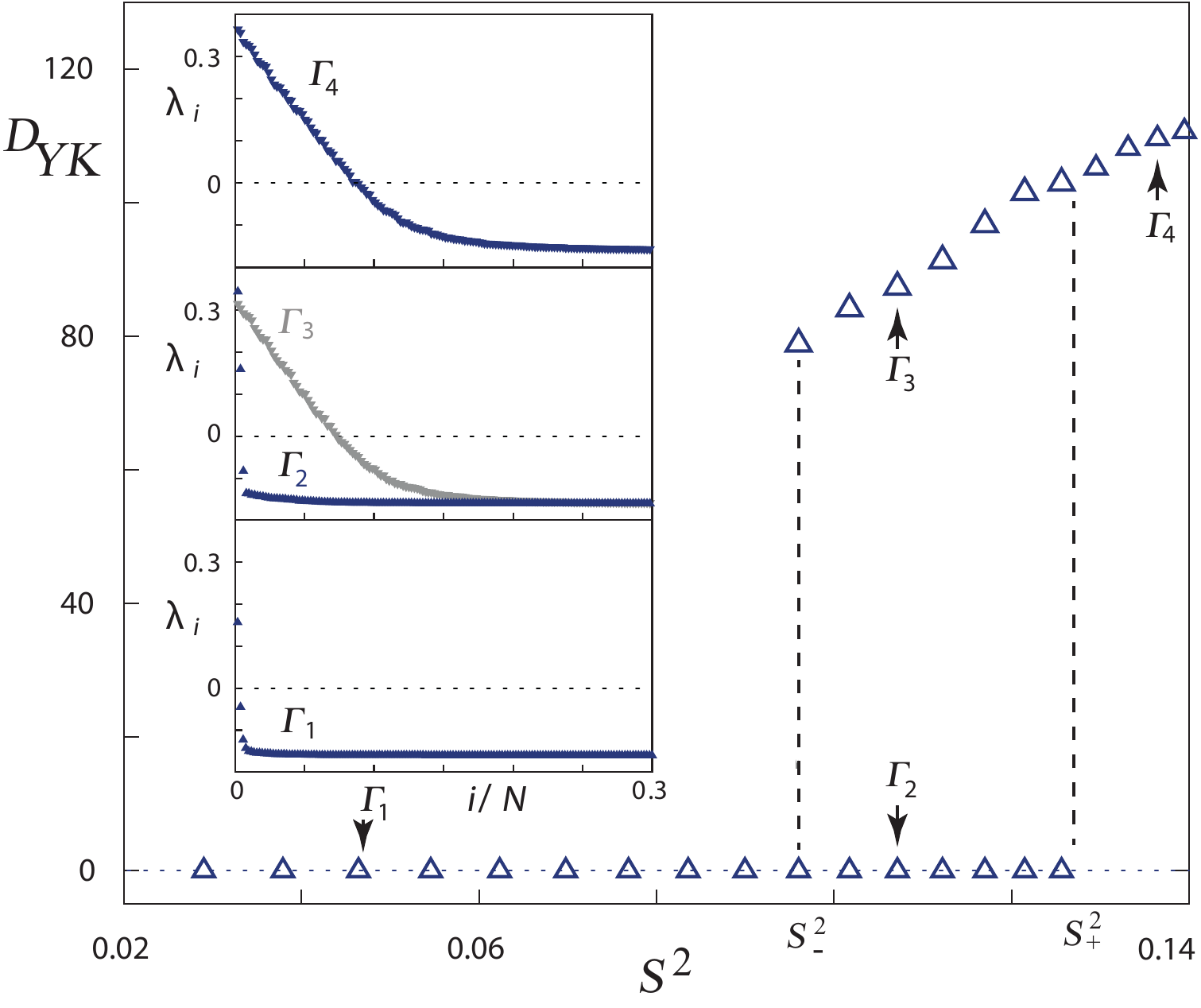} 
\caption{(color online) { Bifurcation diagram of spatiotemporal chaos showing the
Yorke-Kaplan dimension, $D_{YK}$, as function of the intensity of pumping  obtained by numerical simulations of  
Eq.~(\ref{Eq-DDNLS}).
The insets account for the  Lyapunov spectra obtained
for four values of the pumping intensity indicated by the symbol $\Gamma_j$ ($j=1,2,3,4$). The parameters are  $\delta=1$,  and
$\alpha=0.16$. The grid points is 512. The spectra are composed by $N=496$ exponents.}} 
\label{Fig-Lyap_spect_and_YK} 
\end{figure}

The model (\ref{Eq-DDNLS}) supports 
stationary localized \cite{Scorggie_csf94} and self-pulsating localized \cite{Vladimirov} structures. 
In the conservative limit, $(\alpha,S) \to (0,0)$, 
 localized structures have analytical solutions \cite{Nozaki1985,Nozaki1986,Taki1989,Wabnitz93}. 
 It has been also shown that in this limit, localized structures can exhibit
regular time oscillations and display a complex behavior \cite{Nozaki1985,Nozaki1986,Taki1989}. 
An {example} of  complex spatiotemporal  behavior is plotted in the $\tau$-$t$ map of Fig.~\ref{Fig-Cavity_figure}{ (a)}. 
The time evolution of the field amplitude that circulates inside the cavity exhibits large amplitude localized pulses. 
{ These pulses have irregular distribution  along the $\tau$ coordinate  (cf. Fig.~\ref{Fig-Cavity_figure}(a))}. The characterization of this behavior can be achieved by means of Lyapunov exponents, 
which provide an information about the sensitivity of close initial conditions  \cite{Pikovsky}. 
When the largest Lyapunov exponent is positive, the system develops chaos but  not necessarily a 
spatiotemporal chaos. To distinguish between these two complex dynamical behavior, it is necessary to
compute the Lyapunov spectra composed by a set of  exponents \cite{Ruelle,Pikovsky,Manneville90}. 
Spatiotemporal chaos has a Lyapunov spectrum with a {\em continuous} set of  positive values. In contrast, chaos possesses  
a Lyapunov spectrum with discrete set of positive values.  The Lyapunov exponents is denoted by 
$\{\lambda_i\}$, where $i$ labels the exponents 
($i = 1, . . . ,N$) and $\lambda_p\leq \lambda_q $ ($p\geq q$). By using the strategy proposed  
in~\cite{Lyapunov}, we compute numerically the Lyapunov spectrum  for large $N$.
The numerical simulations are obtained by using periodic boundary conditions that are compatible  with both  
Kerr optical frequency combs and  fiber cavities geometries.  
Figure~\ref{Fig-Cavity_figure}(b) show{s} a typical continuous Lyapunov spectrum.
Hence, we  infer that the complex  dynamical behavior
shown in Fig.~\ref{Fig-Cavity_figure}(a)  is a spatiotemporal chaos.

The main feature of the Lyapunov spectra is that they are proportional to the physical 
system size. This implies that the upper limit of the strange attractor dimension  of spatiotemporal 
chaos---the Kaplan-Yorke dimension ($D_{YK}$)---is  an extensive quantity 
that increases with the physical system size \cite{Ruelle}. 
This latter quantity provides an information on the level of 
 the strange attractor complexity and is defined by~\cite{Ott2002} 
\begin{equation}
D_{YK} \equiv p+\frac{\sum_{i=1}^{p}\lambda_{i}}{\lambda_{p+1}},
\end{equation}
where $p$ is the largest integer that satisfies $\sum_{i=1}^{p}\lambda_{i}>0$.
Figure~\ref{Fig-Cavity_figure}(c) displays $D_{YK}$ as function of 
 the number of discretization points, 
which shows that this dimension is indeed an extensive physical quantity  as it linearly increase{s}  with the system size. 
Therefore, as one increases the {system size},  
the dimension of the strange attractor grows proportionally.

To establish the bifurcation diagram of the spatiotemporal chaos,
we fix the detuning  and the dissipation values and { we numerically estimate $D_{YK}$ by varying the pumping intensity}. The initial condition consists of a single peak localized structure. 
The summary of  the results is illustrated  in figure~\ref{Fig-Lyap_spect_and_YK}.  
When increasing pump intensity, the LLE has a zero York-Kaplan dimension, i.e., $D_{YK}=0$  until the system reaches $S^2 \equiv S_+^2$. For $S^2 > S_+^2$, the system exhibits a transition towards a spatiotemporal chaos, i.e., $D_{YK}>0$. This behavior { lasts} for large pumping intensity values. When decreasing $S^2$, the spatiotemporal chaos { persists down} to the point  $S^2 \equiv S_-^2$ as shown in    figure~\ref{Fig-Lyap_spect_and_YK}. From this figure, we clearly see an hysteresis loop involving a spatiotemporal chaos,
a pulsating localized structure and  a HSS  in the range  $S_-^2<S^2<S_+^2$. The inset in Fig.~\ref{Fig-Lyap_spect_and_YK} shows the  continuous Lyapunov spectra for different values 
 of the pump intensity.
Remarkably, the middle panel of the inset shows two Lyapunov spectra $(\Gamma_2$ and $ \Gamma_3)$ 
obtained for the same parameters values indicating the coexistence of two qualitatively  different dynamical behaviors.

\begin{figure}[t]
\centering 
\includegraphics[width=8.5 cm] {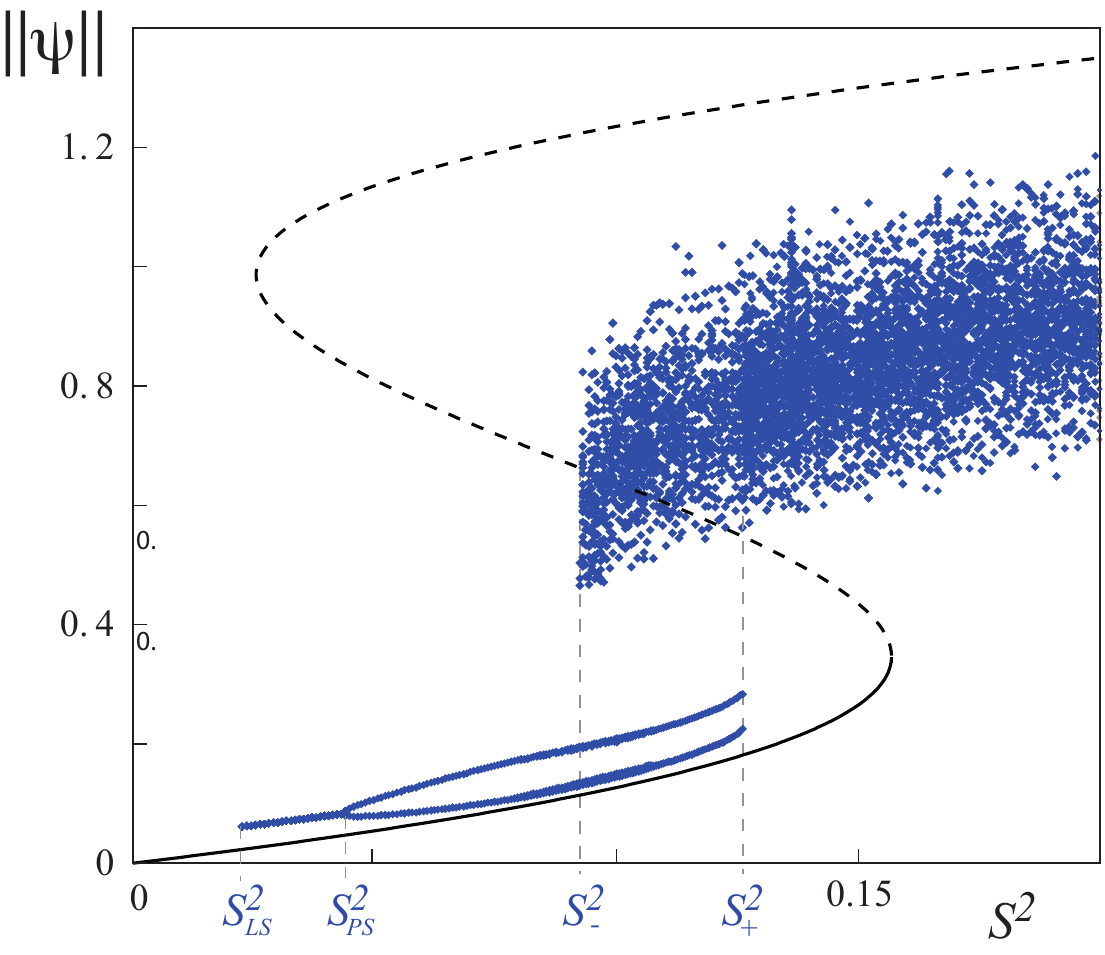} 
\caption{  (color online) Bifurcation diagram of model Eq.~(\ref{Eq-DDNLS}).
The  total intracavity  intensity {$||\psi||$ versus the pump intensity 
$S^2$ with $\delta=1$, and $\alpha=0.16$. The  continuous and dashed thick 
line points out the stable and unstable HSS, respectively.  The continuous blue lines 
indicate the extrema of the total intracavity  intensity $||\psi||$ of localized states}. 
The cloud of blue scattered points accounts for the spatiotemporal chaotic state.} 
\label{Fig-Bistability_curve} 
\end{figure}

\begin{figure}[t]
\centering 
\includegraphics[width=6.0 cm] {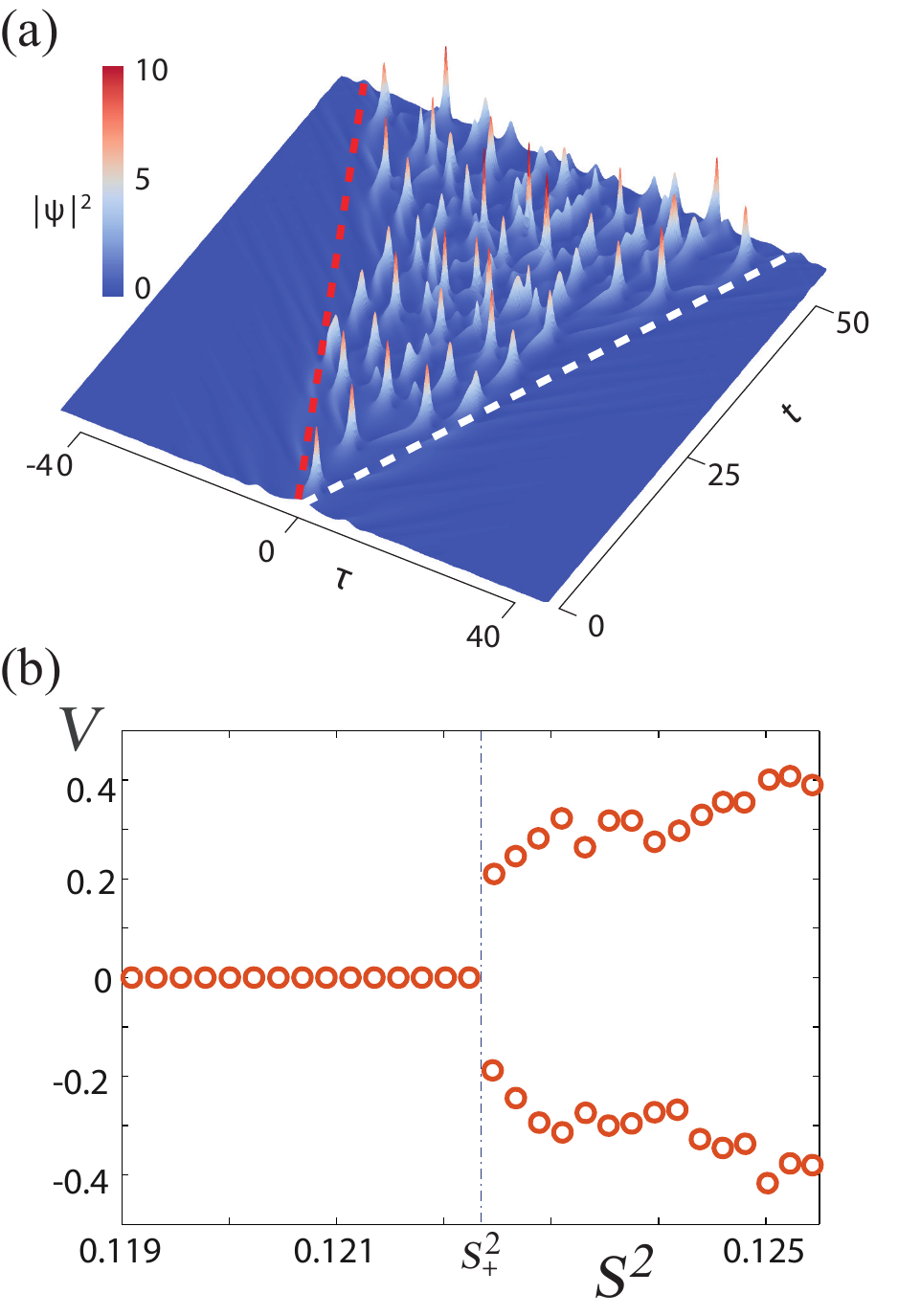} 
\caption{ (color online)  Fronts radiation from an oscillating unstable localized state.
(a) Spatiotemporal evolution of oscillatory localized structures 
obtained from numerical simulation of Eq.~(\ref{Eq-DDNLS}). The parameters are 
$S=0.35$,  $\delta=1$, and $\alpha=0.16$. The dashed lines mark separation between the chaotic and the  homogeneous background.
From these lines one can determine the front speed.
(b) The front speed $V$ as a function of the pump intensity obtained for $\delta=1$, and $\alpha=0.16$.} 
\label{Fig-FrontPropagation} 
\end{figure}

In what follows, we establish a bifurcation diagram showing a coexistence between the spatiotemporal 
chaos, the oscillatory localized structure, and the HSS. In order to show 
different operating regimes, the total  intracavity field amplitude 
$||\psi||\equiv \int |\psi(t,\tau)|^2 d\tau$ 
as a function of the pumping 
intensity is shown in the bifurcation diagram figure~\ref{Fig-Bistability_curve}. The upper (lower) HSS branch indicated by dashed (solid) line is modulationally 
unstable (stable) \cite{Lugiato1987}. 
For small pumping intensity, the system has stationary stable localized state in the range $S_{LS}^2<S^2<S_{PS}^2$ 
(see Fig.~\ref{Fig-Bistability_curve}). When increasing the pumping intensity  the localized state becomes self-pulsating
in the range $S_{PS}^2\leq S^2<S_{+}^2$. When further increasing $S^2$, the system exhibits spatiotemporal chaos.
When decreasing $S^2$, the spatiotemporal chaos persist{ s down} to $S_{-}^2$. As in the bifurcation diagram of $D_{YK}$
(cf. fig.~\ref{Fig-Lyap_spect_and_YK}), the system presents an hysteresis loop involving three different robust states:
HSS, pulsating localized structures, and spatiotemporal chaos.
 
 It is well known that model~(\ref{Eq-DDNLS}) exhibits  
radiation from a localized state of two counter-propagative fronts between the homogeneous and the complex  
spatiotemporal states \cite{LeoCoen}. 
An example of this behavior is depicted in the $\tau$-$t$ map shown in figure~\ref{Fig-FrontPropagation}(a). 
To characterize this transition,  we estimate numerically the front speed. Figure~\ref{Fig-FrontPropagation}(b) shows the front speed as a function of the pump intensity  in the vicinity of the instability associated with localized states. 
Right and left fronts propagate with almost the same speed. As the pumping intensity is increased, 
the front speed continues to increase until the system reaches the lower limit point of  bistable  HSSs. 
Similar behavior { has} been reported in pattern forming 
systems where front propagates between a HSS and a periodic pattern \cite{Pomeau,ClercFalcon,ClercFalcon2} or between either two HSSs \cite{Coulibaly1,Coulibaly2} or even between a HSS and the spatiotemporal intermittency \cite{Michel2016}. 
 
From practical point of view, a  driven ring cavity made with an optical fiber could support a spatiotemporal regime. 
However,  by using  constant injected beam, i.e., cw operation, it is hard to reach the high intensity  
regime where we can observe 
the  spatiotemporal chaos and its coexistence with a homogeneous background.  To overcome this limitation, it is necessary to  
drive the cavity with { a synchronously pumping with a pulsed laser}. The time-of-flight of the light pulses in the cavity should be
adjusted to the laser repetition time. All experiments using this simple device with pulse laser have shown 
evidence of complex  spatiotemporal behaviors \cite{Mitschke96,Mitschke96b,Coen_Optica_15,Mitschke98}. Therefore,
the phenomenon described in this letter should be observed experimentally. 

In conclusion, by using rigorous tools of dynamical system theory, such as Lyapunov spectra, 
we have  quantitatively shown  that the complex behavior observed experimentally in the Kerr 
optical frequency combs \cite{Chembo_10}  and in the fiber cavity \cite{Mitschke96,Mitschke96b,Coen_Optica_15} 
is of a spatiotemporal chaos nature. 
We have also shown that the Yorke-Kaplan dimension  
can be considered as a good order-parameter 
to characterize the bifurcation diagram associated with spatiotemporal chaos. Finally, we have identified different operating 
regimes, in particular the coexistence between spatiotemporal chaos, the self-pulsating localized 
structure, and the homogeneous steady state.
The observed complex states are exponentially sensitive to the initial conditions, exhibit  
complex spatiotemporal chaos, and have  exponential power spectrum. Hence, this  behavior is not of turbulent nature. 
Our finding is therefore important for the analysis, classification of the various complex 
spatiotemporal behaviors observed in practical dissipative systems.
 
 M.G.C. thanks for the financial support of FONDECYT projects 1150507. Z.L, S.C and M.T thank the Interuniversity Attraction Poles program of the Belgian Science Policy Office under the grant IAPP7-35, the French Project ANR Blanc OptiRoc N12-BS04-0011, and the "Laboratoire d'Excellence: Centre Europ\'een pour les Math\'ematiques, la Physique et leurs Interactions" CEMPI. M.Tlidi received support from the Fonds National de la Recherche Scientifique (Belgium).

\end{document}